\documentclass[conference]{IEEEtran}

\usepackage{lineno}
 
\usepackage{booktabs} 
\usepackage{balance} 
\usepackage{tabularx}
\usepackage{lipsum}

\usepackage{amsmath,amssymb,amsfonts}
\providecommand{\keywords}[1]{\textbf{\textit{Index terms-}} #1}
\usepackage{graphicx}
\usepackage{subcaption}
\usepackage{textcomp}
\usepackage{fleqn}
\usepackage{soul}
\usepackage{color}
\usepackage{subcaption}
\usepackage{caption}
\usepackage{floatrow}%
\usepackage{wrapfig}
\usepackage{comment}
\usepackage{relsize}
\usepackage{algorithm}
\usepackage{algpseudocode} 
\usepackage{pifont}
\usepackage{tikz}
\usepackage{float}
\usepackage{amssymb}
\usepackage{floatrow}
\usepackage{cite}
\usepackage{xcolor}
\usepackage{lipsum}
\usepackage{enumitem}

\modulolinenumbers[5]
\usepackage{pdflscape}

\begin{document}

\title{Spatial-Interference Aware Cooperative Resource Allocation for 5G NR Sidelink Communications}
\author{\IEEEauthorblockA{ Silvia Mura\IEEEauthorrefmark{1}, Francesco Linsalata\IEEEauthorrefmark{1}, Marouan Mizmizi\IEEEauthorrefmark{1}, Maurizio Magarini\IEEEauthorrefmark{1}, \\ Majid Nasiri Khormuji\IEEEauthorrefmark{2}, Peng Wang\IEEEauthorrefmark{2}, Alberto Perotti\IEEEauthorrefmark{2}, Umberto Spagnolini\IEEEauthorrefmark{1}\IEEEauthorrefmark{3}}

\IEEEauthorblockA{\IEEEauthorrefmark{1}Dipartimento di Elettronica, Informazione e Bioingegneria, \\ Politecnico di Milano, Via Ponzio 34/5, 20133, Milano
Italy} 
\IEEEauthorblockA{\IEEEauthorrefmark{2}Huawei Technologies Sweden AB, Skalholtsgatan 9-11, SE-164 94 Kista, Stockholm, Sweden}
\IEEEauthorblockA{\IEEEauthorrefmark{3}Huawei Industry Chair} 
E-mails: \{silvia.mura, francesco.linsalata, marouan.mizmizi\}@polimi.it}
\maketitle

\maketitle

\begin{abstract} 
Distributed resource allocation (RA) schemes have been introduced in cellular vehicle-to-everything (C-V2X) standard for vehicle-to-vehicle (V2V) sidelink (SL) communications to share the limited spectrum (sub-6GHz) efficiently. However, the recent progress in connected and automated vehicles and mobility services requires a huge amount of available spectrum resources. Therefore, millimeter-wave and sub-THz frequencies are being considered as they offer a large free bandwidth. However, they require beamforming techniques to compensate for the higher path loss attenuation. The current fifth-generation (5G) RA standard for SL communication is inherited from the previous C-V2X standard, which is not suited for beam-based communication since it does not explore the spatial dimension. In this context, we propose a novel RA scheme that addresses the directional component by adding this third spatial dimension to the bandwidth part structure and promotes cooperation between vehicles in resource selection, namely cooperative three-dimensional RA. Numerical results show an average of 10\% improvement in packet delivery ratio, an average 50\% decrease in collision probability, and 30\% better channel busy ratio compared to the current standard, thus, confirming the validity of the proposed method.
\end{abstract}


\section{Introduction}

In recent years, the advances in vehicle-to-everything (V2X) communications have boosted the evolution of intelligent transportation systems (ITS) toward a safer and more efficient road network. This has also led to advanced services with increasingly stringent communication requirements. Currently deployed V2X technologies, such as dedicated short-range communication (DSRC)~\cite{dsrc} and long-term evolution (LTE) cellular-V2X (C-V2X)~\cite{Sepulcre8691973}, operate at sub-6GHz frequencies. Compared to DSRC, C-V2X exhibits better performances thanks to resource allocation (RA) schemes that allow for efficient sharing of the available bandwidth while reducing mutual interference~\cite{Sepulcre8581518}.

In the context of LTE C-V2X, there are two classes of RA schemes: \textit{(i)} \textit{centralized RA}, the base station allocates the resources for all the vehicles in the network, e.g., mode 3, and \textit{(ii)} \textit{distributed RA}, the vehicles autonomously sense and select the available resources, i.e., mode 4~\cite{9127428}. The authors in~\cite{he2020interference} propose a short-term sensing window for periodic and aperiodic traffic for C-V2X, where sidelink (SL) beam-based communications are not considered. The work in~\cite{molina2019geo} proposes a resource selection based on the bandwidth partitioning according to the CAVs' positions. However, this approach can limit the performances by encouraging collisions in frequency resources selection. 
The same drawback can affect the RA solution proposed in~\cite{kim2018position}, where frequency resources are allocated based on vehicles' speed, density, heading, and positions. Thus, the static partitioning of the bandwidth according to the scenario topology may affect the candidate sub-channels selection. Cluster-based solutions are used to deal with semi-distributed RA. These involve the election of a cluster head that manages resource sharing and power allocation. In~\cite{zhao2019cluster} the bandwidth is partitioned dynamically according to the cluster definition, but the presence of an elected vehicle may cause undesired latency.

The literature mainly focuses on sub-6GHz portion of the spectrum with a limited bandwidth that barely meets the requirements of basic V2X services. To accommodate the needs of enhanced V2X (e-V2X) services, the third generation partnership project (3GPP) has proposed the use of millimeter-wave (mmW) bands for fifth-generation (5G) NR-based V2X systems~\cite{6g}, and sub-THz frequencies are being investigated for the upcoming sixth-generation (6G) V2X communication.  
Wireless propagation at such very high frequencies is subject to an order of magnitude higher path loss attenuation and needs beamforming techniques to ensure the required coverage~\cite{Rappaport:J13}. The centralized mode 1 and distributed mode 2 schemes for RA in 5G NR are inherited from the C-V2X mode 3 and mode 4, respectively. However, these fail to properly include the spatial dimension introduced by the beam-type communication. Therefore, the resulting interference, in particular for vehicle-to-vehicle (V2V) SL communication, can compromise the 5G NR RA effectiveness and it results in poor reliability~\cite{Petrov_inter}.
\begin{figure*} [t!]
    \centering
    \includegraphics[width=0.65\textwidth]{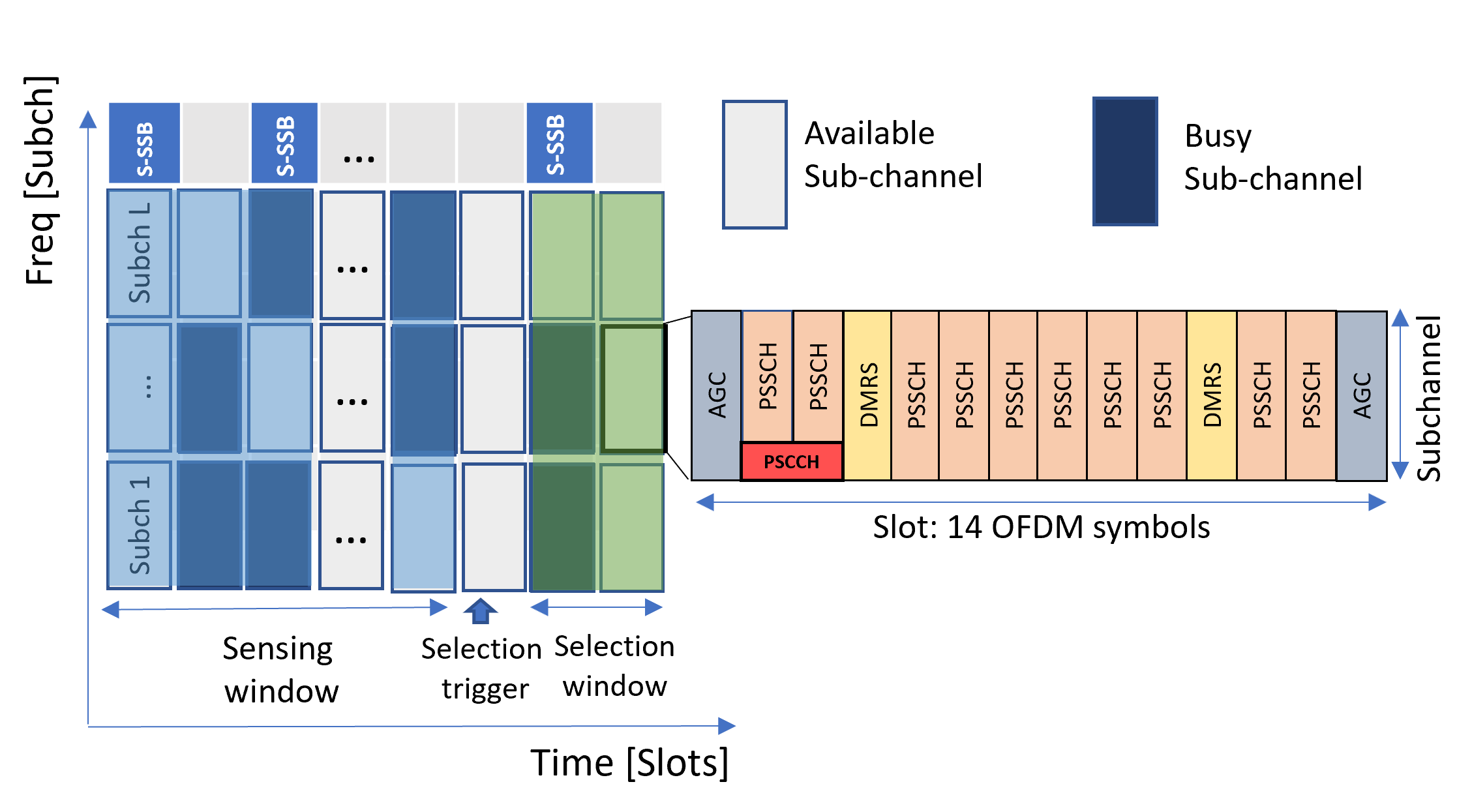}
    \caption{Bandwidth part (BWP) for sidelink RA representing the sensing and selection window and sub-channel structure.}
    \label{fig:BWPscheme}
\end{figure*}
As main result of this paper, we show that the standard RA procedure is not properly suited to the new 5G physical layer (PHY) paradigm. Therefore, we propose a novel three-dimensional (3D) cooperative RA scheme that accounts for the directional component and, when compared with the current standard RA, it improves the packet delivery ratio (PDR) by 10\% on average, the channel busy ratio (CBR) of the bandwidth of 30\%, and halves the average collision probability.

The remainder of this paper is organized as follows. The PHY design of SL 5G NR is discussed in Sec.~II. Section~III presents the semi-persistent RA and the cooperative 3D RA schemes. Simulation setup and numerical results are discussed in Sec.~IV. Finally, Sec.~V draws the conclusions.

\section{Sidelink 5G NR PHY}

This section highlights the main features of the current 5G NR PHY structure of SL RA, such as the bandwidth part (BWP) structure (Sec.~\ref{sec:bwp}), the logical channels (Sec.~\ref{sec:log}), and the RA modes (Sec.~\ref{sec:modes}).
\subsection{Bandwidth Part}\label{sec:bwp}
The BWP, depicted in Fig.\ref{fig:BWPscheme}, defines the 2D resource grid in the time and frequency domains. SL transmissions and receptions are contained within the SL BWP. 
The minimum BWP element, called resource block (RB), spans one-time slot, composed of 14 OFDM symbols, and 12 consecutive sub-carriers with the same sub-carrier spacing (SCS). The BWP numerology $\mu$ is related the slot duration as $2^{-\mu}$ ms and to the SCS, evaluated as $2^{\mu} \times$ 15 kHz~\cite{TS_38214}. The sub-channel represents the smallest unit for a SL data transmission or reception and it is composed of a pre-configured number of consecutive RBs. 
The ensemble of sub-channels dedicated to SL data transmission defines the resource pool (RP) and it does not include the resources allocated for the initial access and synchronization~\cite{sidelink, LinsICC-c21}.

\subsection{Logical Channels}\label{sec:log}

The 5G NR SL provides the following logical channels~\cite{phychanmod}

\begin{itemize}[wide]
    \item \textbf{Physical Sidelink Control Channel (PSCCH):} carries the 1st-stage sidelink control information (SCI), such as the information regarding the PSCCH sub-channels resources, the information for decoding the 2nd-stage SCI, and the priority of the transport block (TB) transmission.
    When a transmitting vehicle (TxV) sends the TB to a receiving one (RxV), it first sends the PSCCH that is decoded for the channel sensing purpose.
      
    \item \textbf{Physical Sidelink Shared Channel (PSSCH):} carries the 2nd-stage SCI and the TB data payload. The 2nd-stage SCI contains the remaining information to decode the PSSCH and the TxV/RxV identifiers (IDs). 
      
    \item \textbf{Physical Sidelink Feedback Channel (PSFCH):} carries the hybrid automatic repeat request (HARQ) feedback concerning the TB reception.
\end{itemize}

\subsection{RA Modes}\label{sec:modes}

SL RA in the current standard defines two RA modes.

\begin{itemize}[wide]
    \item \textbf{Centralized Mode 1:} the SL resources are allocated by the base station when vehicles are in its coverage area. It can use either (i) dynamic grant scheduling, i.e., for each TB, the base station allocates the dedicated resources, or (ii) configured grant scheduling, which enables the reservation of sub-channels for transmitting multiple TBs~\cite{mode1}.
    
    \item \textbf{Distributed Mode 2:} it is used when vehicles are out of coverage and they autonomously select SL resources from the RP. It can operate either in the dynamic scheduling, where new resources are selected for the transmission of each TB or in the semi-persistent scheduling, where resources are reserved for the transmission of multiple TBs~\cite{9345798}.
\end{itemize}

The semi-persistent scheduling is more suited to latency-critical applications than the dynamic one since the resources are allocated for multiple TBs~\cite{garcia2021tutorial}.
Due to its generality, in the next section, we present the 5G NR standard semi-persistent scheme and we discuss its limitation in the mmW V2V communications.

\section{Sidelink Semi-Persistent Scheme}

The semi-persistent RA scheme is pre-configured in the RP, and each vehicle, after the initial access and synchronization, can select the resources for the transmission of several TBs. The maximum number of consecutive TBs that can be allocated is defined by the \textit{re-selection counter} (RC), and it depends on the \textit{resource reservation interval}(RRI), which defines the periodicity for the allocated TBs. 
These semi-persistent parameters are mutually related: if $RRI$ $\geq$ $100\,$ms RC is randomly selected in $[5,15]$ , otherwise the counter is randomly set within the interval $[5C,15C]$ with $C $\,=\,$ $100\,/$\text{max}(20,RRI)$ and it is decremented by 1 after each transmission~\cite{8613007}. When the RC value is equal to zero, new resources are selected with probability $1-P$, with \textit{P} $\in [0, 0.8]$~\cite{bazzi2018study}. In the following, the standard SL RA scheme is detailed.

\subsection{Standard-compliant RA}

The standard SL RA scheme consists of two phases~\cite{15rel}, the \textit{sensing} of the available resources, followed by the \textit{selection} of the transmission time-frequency resources.

\subsubsection{Sensing}

each vehicle, when not transmitting, senses for a time interval defined by the sensing window for the available resources in the RP by measuring the received signal reference power (RSRP) for each sub-channel. If RSRP $\leq \gamma_{th}$, the sub-channel is available, otherwise busy.  The threshold $\gamma_{th}$ depends on the priority of the TBs to be transmitted and the percentage of occupied RBs in the sensing window. Therefore, the $i$th vehicle stores a bitmap matrix $\mathbf{B}_i$ defined as 
\begin{align} \label{eq:bitmap}
    \left[\mathbf{B}_i\right]_{(t,f)} =
    \begin{cases}
        1,               & \text{if } \text{RSRP}\leq \gamma_{th},\\
        0,               & \text{otherwise},
    \end{cases}
\end{align}
with $f$ and $t$ referring to the sub-channel and time-slot indexes, respectively. The ratio between the number of candidate resources and the total number of resources in the sensing window is calculated by each vehicle as 
\begin{align}\label{eq:perc}
p = \frac{|\mathbf{B}_i|_{0} \cdot 100}{|\mathbf{B}_i|},
\end{align}
where $|.|_{0}$ refers to the $l_{0}$-norm and $|.|$ to the matrix cardinality. 
In highly congested networks, $p$ is low, which implies that limited resources are available. Thus, when $p \leq \Gamma_{th}$, the threshold $\gamma_{th}$ is decreased by $3\,$dB, and the above procedure is repeated.
When RSRP $> \gamma_{th}$, the TxV attempts to decode the PSCCH and the related 1st-stage SCI, which contains the RRI information. The RRI is exploited to estimate the RC, and by combining RRI and RC, the TxV predicts the resources that have been allocated and excludes them from the available resources.
Figure~\ref{fig:BWPscheme} depicts the SL-BWP perceived by a vehicle; in the sensing window each sub-channel is categorized as \textit{busy} or \textit{available}, according to measured RSRP.

\subsubsection{Selection}

when a TB is ready for transmission, a selection trigger is activated. Among the available resources, the vehicle chooses $M$ candidates for TB transmission and $M-1$ candidates for TB re-transmission or HARQs~\cite{TS_38213}, with $1 \leq M \leq 32$. Then, the vehicle randomly selects the resources among the candidate ones. 
Re-selection and re-assessment are introduced by 3GPP to manage traffic priority and TB latency requirements. However, these are still under discussion for further enhancements of the congestion control in Rel.~17 and they are not included in this work.

The pseudocode of the standard semi-persistent RA is reported in Alg.~\ref{alg:Semipersistent}.

\begin{algorithm}[!b] 

\caption{Standard-complaint RA}
\label{alg:Semipersistent}
\begin{small}
\begin{algorithmic}[1]
\Statex{\textbf{Sensing}}
\State\,\, RSRP measurement for each sub-channel $f$ 
\State\,\, Update bitmap $\mathbf{B}_{i}$ according to \eqref{eq:bitmap}
\State\,\, Evaluate $p$ as in \eqref{eq:perc}
\Statex\,\, \textbf{if} $p > \Gamma_{th}$ 
    \State\,\, \hspace{0.3cm}{ RRI of the sub-channels (PSCCH) decoding} 
    \State\,\, \hspace{0.3cm}{ Estimate RC}
    \State\,\, \hspace{0.3cm}{$\left[\mathbf{B}_i\right]_{(t_{RRI},f)} $\,=\,$ 0$}
    \vspace{0.1cm}
    \Statex\,\, {\hspace{0.9cm}with $t_{RRI}$ $\,=\,$ $\{t, t+RRI,..., t+(RC-1)\, RRI\}$ }
\Statex\,\, \textbf{else}
    \State\,\,{\hspace{0.3cm}$\gamma_{th} $\,=\,$ \gamma_{th} -3$ [dB]}
    \State\,\,{\hspace{0.3cm}Repeat from step 1}
\Statex\,\, \textbf{end}
\Statex{\textbf{Selection}}
\State {TxV randomly selects sub-channels and RRI based on $\mathbf{B}_i$}
\end{algorithmic}
\end{small}
\end{algorithm}
\begin{figure} [t!]
    \centering
    \includegraphics[width=0.95\textwidth]{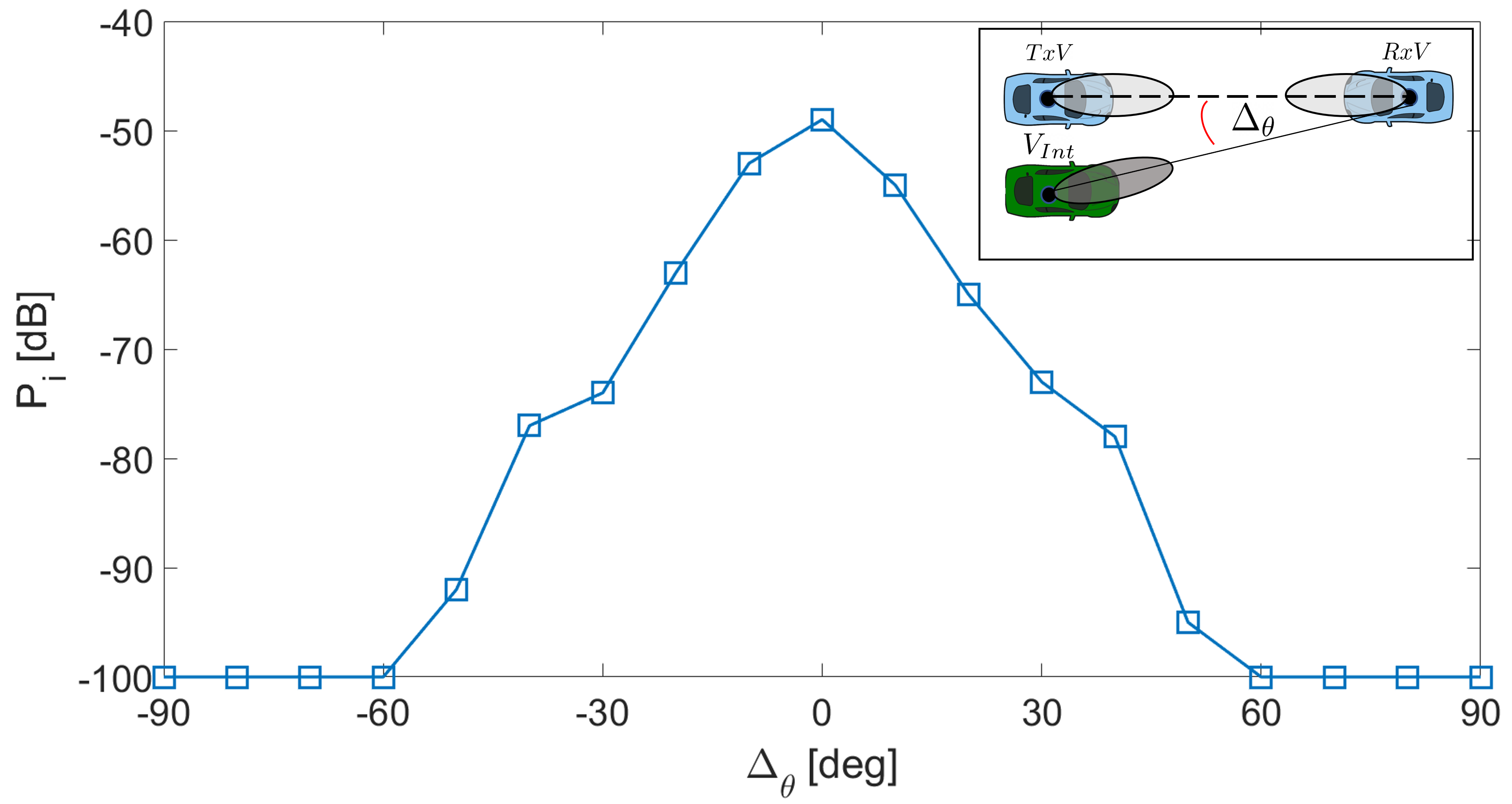}
    \caption{Simulated Interference power $P_{i}$ versus $\Delta_{\theta}$.}
    \label{fig:angleInterf}
\end{figure}
\subsection{Cooperative 3D RA}
\begin{figure} [t!]
    \centering
    \includegraphics[width=0.8\textwidth]{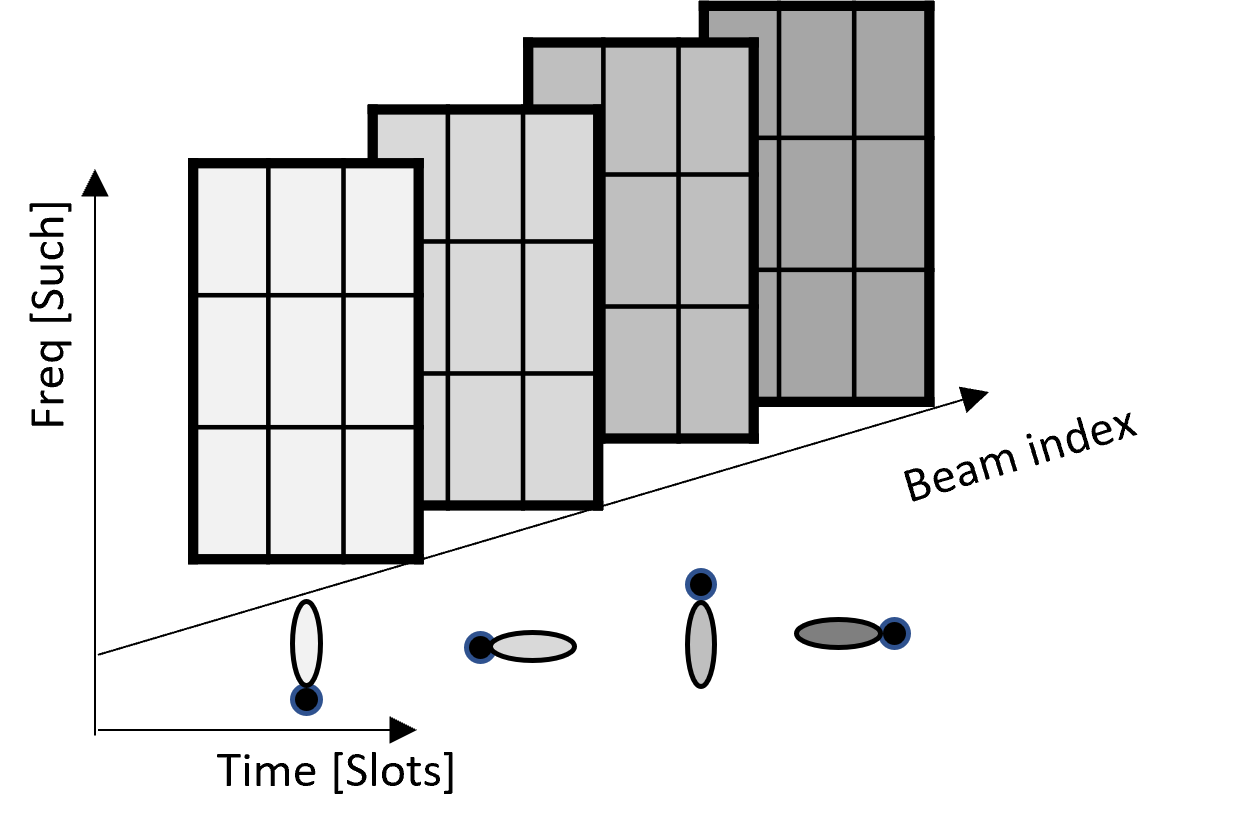}
    \caption{3D resource pool for sidelink RA with beam pointing directions.}
    \label{fig:3DBWPscheme}
\end{figure}

\begin{algorithm}[!t] 
\caption{Cooperative 3D RA}
\label{alg:3D}
\begin{small}
\begin{algorithmic}[1]
\Statex{\textbf{Sensing}}
\Statex \,\,\textbf{for} $\theta \in \boldsymbol{\Theta}$ 
    \State\,\,\,\,{\hspace{0.3cm}RSRP measurement for each sub-channel $f$}
    \State\,\,\,\,{\hspace{0.3cm}Update bitmap $\mathbf{B}_{i,\theta}$ according to (\ref{eq:bitmap})}
    \State\,\,\,\,{\hspace{0.3cm}Evaluate $p_{\theta}$ as in \eqref{eq:perc}}
    \Statex \,\,\hspace{0.3cm}\textbf{if} $p_{\theta} > \Gamma_{th}$ 
        \State\,\,  {\hspace{0.6cm} RRI of the sub-channels (PSCCH) decoding} 
        \State\,\,  {\hspace{0.6cm} Estimate RC}
        \State\,\,  {\hspace{0.6cm} $\left[\mathbf{B}_{i,\theta}\right]_{(t_{RRI},f)} $\,=\,$ 0$}
        \vspace{0.1cm}
        \Statex\,\, {\hspace{1.2cm}with $t_{RRI}$\,=\,$\{t, t+RRI,..., t+(RC-1) RRI\}$ }
    \Statex \,\,\hspace{0.3cm}\textbf{else}
        \State\,\, {\hspace{0.6cm}$\gamma_{th} $\,=\,$ \gamma_{th} -3$ [dB]}
        \State\,\, {\hspace{0.6cm}Repeat from step 1}
    \Statex \,\,\hspace{0.3cm}\textbf{end}
\Statex\,\, \textbf{end}
\Statex{\textbf{Selection}}
\State {TxV sends RA request to RxV}
\State{RxV randomly selects the sub-channels and RRI based on $\mathbf{B}_{i,\bar{\theta}}$}
\State {RxV feedbacks the RA configuration to TxV}
\end{algorithmic}
\end{small}
\end{algorithm}

The standard procedure for RA suffers from several drawbacks as it is not suited for beam-based mmW communications. The main issues are related to the sensing procedure, in which the spatial interference is not addressed.  
This can be observed in Fig.~\ref{fig:angleInterf}, where the interference power is evaluated varying the angle of the interfering vehicle.
This result shows that the interference, which may affect the perception of available sub-channel resources, is mainly related to the beam pointing directions. Moreover, in the standard RA, the TxV allocates resources based on its sensing, and it may not sense the interference affecting the RxV. 

We propose a RA scheme that explores the spatial dimension given by the beams, and the cooperation among vehicles, i.e., cooperative 3D RA as shown in Fig.~\ref{fig:3DBWPscheme}. Compared to the standard RA, the proposed approach introduces two main novelties: \textit{(i)} during the selection phase, the cooperative 3D RA employs a bitmap $\mathbf{B}_{i,\theta}$ for each beam $\theta$ defined in the beamforming codebook $\Theta $\,=\,$ [\theta_1,\dots,\theta_L]$, where $L$ is the codebook depth and, \textit{(ii)} cooperation between TxV and RxV in the candidate resource selection. In the selection phase, the TxV sends an RA request to the RxV, which replies with the set of resources available according to its $\mathbf{B}_{i,\bar{\theta}}$ bitmap, where $\bar{\theta}$ refers to its receiving spatial direction. The request for RA and the response from the RxV are signaling messages sent over a set of resources selected using the standard RA selection procedure. The cooperative 3D RA procedure is shown in Alg.~\ref{alg:3D}.

\section{Simulation Results}

\begin{table}[b!] 
\centering
\caption{PHY design parameters}
\footnotesize
\begin{tabular}{ | c | c | }
	\hline
	\textbf{Parameter}  & \textbf{Value} \\ \hline\hline
	Numerology ($\mu$) & $3$ \\ 
	Sub-Carrier Spacing (SCS) & $120$ kHz  \\ 
	Frequency Range & FR2 \\ 
	Cyclic Prefix & Normal \\ 
	OFDM symbols/Slot & $14$ \\ 
	Slots/Subframe & $8$ \\
	Slot duration & $0.125$ ms\\
	Maximum carrier bandwidth & $400$ MHz\\
    Sub-channels/Bandwidth part & $10$\\
    Transmitted Power & $23$ dBm \\ 
    Noise Power & $-68$ dBm  \\ 
%    LNA Noise Figure & $13$ dB \\ 
    Carrier Frequency & $30$ GHz \\ 
    Antenna Height (wrt rooftop) & $0.1$ m \\ 
    Antenna Elements & $4\times4$ \\
    TB Size & $1$ slot\\
    PSCCH/PSSCH Modulation & QPSK\\
\hline
	\end{tabular}
\label{tab:BWPparameter}
\end{table}
\begin{table}[!b] 
\centering
\caption{Semi-persistent scheduling parameters}
\footnotesize
\begin{tabular}{ | c | c | }
	\hline
	\textbf{Parameter}  & \textbf{Value} \\ \hline\hline
	 Resource Reservation Interval (RRI) & $2$ ms \\ 
	 Re-selection Counter (RC) & 4  \\ 
     Sub-channels/Resource pool& $10$ \\ 
	 Resource Blocks PSCCH/Sub-channel & $10$ \\ 
	 Resource Blocks PSSCH/Sub-channel & $12$ \\ 
	 PSCCH symbols/Slot & $2$\\
	 DMRS symbols/Slot & $3$\\
	 Sensing Window & $796$ slots\\
	 Selection Window (W) & $20,40$ slots\\
	 RSRP threshold ($\gamma_{th}$) & $-60$ dB\\
	 Resource occupancy threshold ($\Gamma_{th}$) & $20 \%$ \\
	 Blind retransmissions  & $3$ \\
	 Maximum number of CAVs for selection ($N_{max}$) & $10,20$\\
	 CBR interval & $50$ slots\\
	\hline
	\end{tabular}
\label{tab:sched_par}
\end{table}
\begin{figure} [t!]
    \centering
    \includegraphics[width=0.65\textwidth, height=5.5 cm]{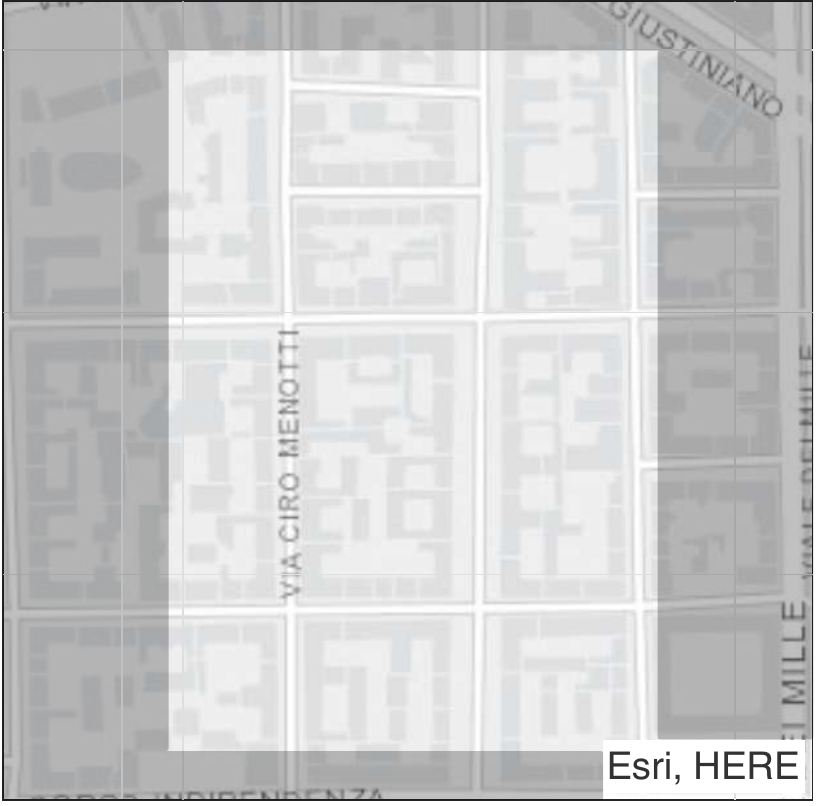}
    \caption{Simulated urban scenario in Milan, Italy lat\,=\,$[45.46848,\,45.47222]$ and lon\,=\,$[9.21239,\, 9.21606]$.}
    \label{fig:scenario}
\end{figure}
Exhaustive numerical simulations were carried out to evaluate the performance of the cooperative 3D RA approach compared with the standard-compliant scheme, considered here as benchmark.
The implemented system model is fully standard-compliant. The latest 3GPP recommendations for the setting of PHY are used~\cite{15rel}. The main PHY parameters are detailed in Tab.~\ref{tab:BWPparameter}, while Tab.~\ref{tab:sched_par} reports those for the 5G NR standard Mode 2 RA.
The communication channel is half-duplex with a bidirectional alternating transmission. Hence, when a vehicle senses the communication channel cannot transmit and vice versa.
Moreover, vehicles employ an array of planar antennas with $4\times4$ configuration, mounted on the rooftop of CAVs as in~\cite{14rel}. The communication between each pair of TxV and RxV is based on the legacy orthogonal frequency division multiple access~\cite{phychanmod}, and the received power in an arbitrary sub-channel is computed as
\begin{align}
    \text{RSRP} = P_t - PL + 2G_b \,\, \text{[dBm]} \,, 
\end{align}
where $P_t$ is the transmitted power, PL is the path loss, and $G_b$ is the beamforming gain~\cite{Perfecto:J2017}. 
The RA algorithms detailed in Algs.~\ref{alg:Semipersistent}~and~\ref{alg:3D}, are applied by vehicles after the initial access and synchronization phase.

\subsection{Simulation Methodology}

The numerical simulations examine an urban area of Milan, Italy, represented in Fig.~\ref{fig:scenario}. The road network scenario, buildings, and trees are obtained from OpenStreeMap~\cite{osm} while the mobility is simulated through Simulator of Urban Mobility (SUMO)~\cite{SUMO2018}. The propagation channels of the SL links are generated based on the latest 3GPP guidelines in \cite{14rel} using a geometric-based efficient propagation model for V2V (GEMV$^2$)~\cite{GEMV2}. This uses the urban topology and the vehicles' mobility to compute the large-scale channel components with a deterministic approach and the small-scale ones with a geometry-based stochastic approach.

\subsection{Performance Metrics}
\begin{figure} [t]
    \centering
    \includegraphics[width=0.8\textwidth, height=5 cm]{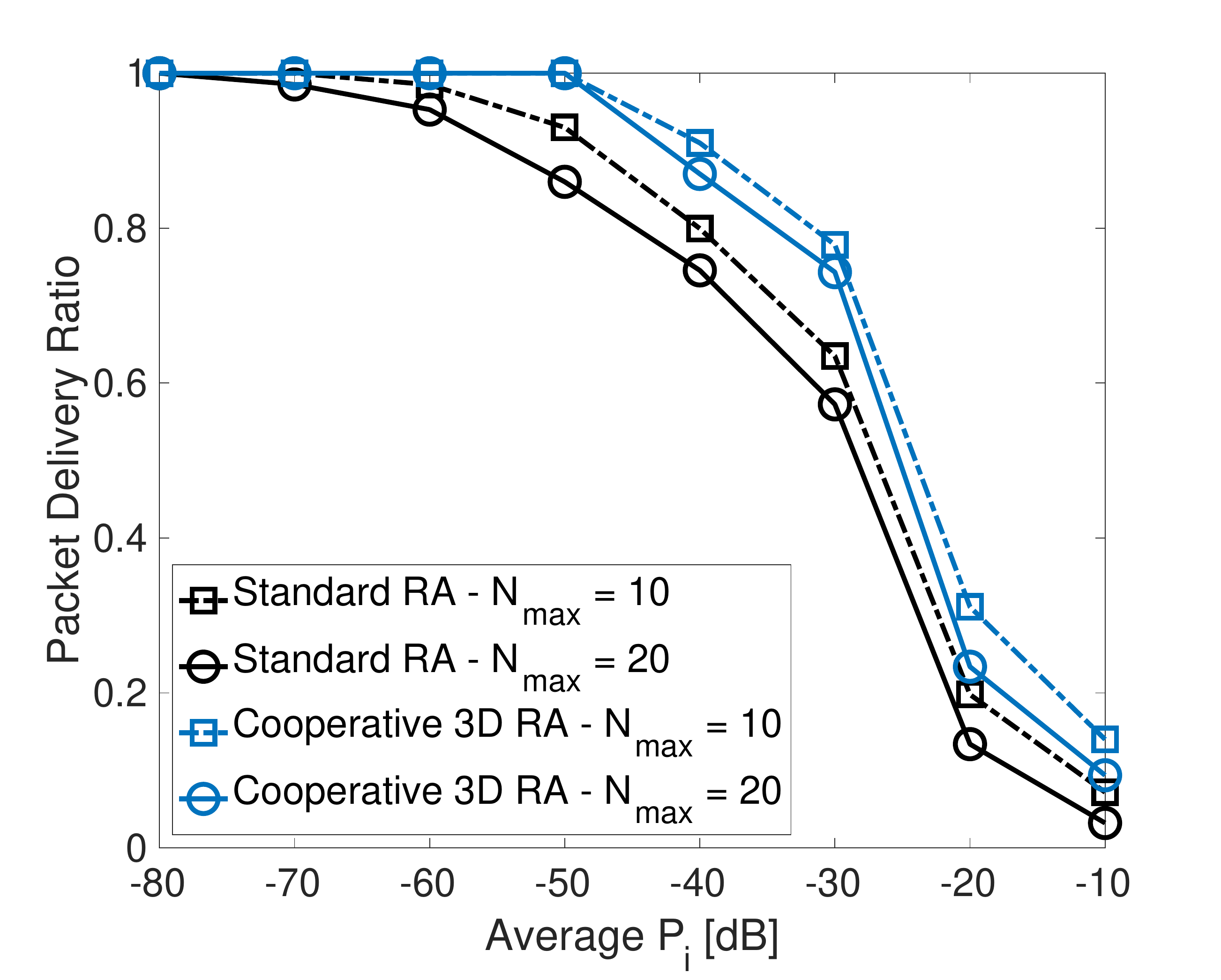}
    \caption{PDR versus average interference power for the two RA schemes.}
    \label{fig:PDR}
\end{figure}
\begin{figure} [t]
    \centering
    \includegraphics[width=0.8\textwidth, height=5 cm]{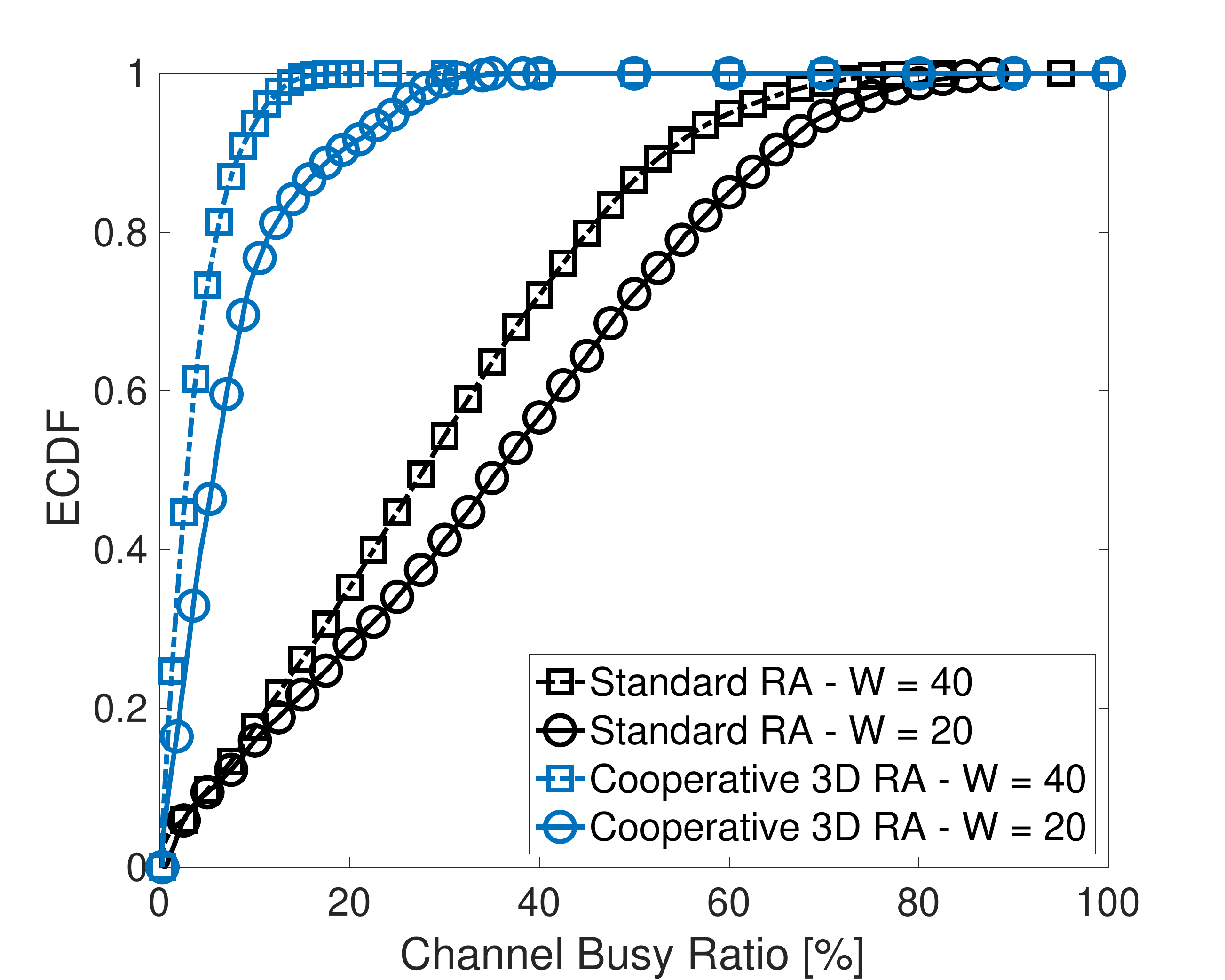}
    \caption{ECDF of the CBR of the two RA approaches.}
    \label{fig:CBR}
    \end{figure}
\begin{figure} [t]
    \centering
    \includegraphics[width=0.8\textwidth, height=5 cm]{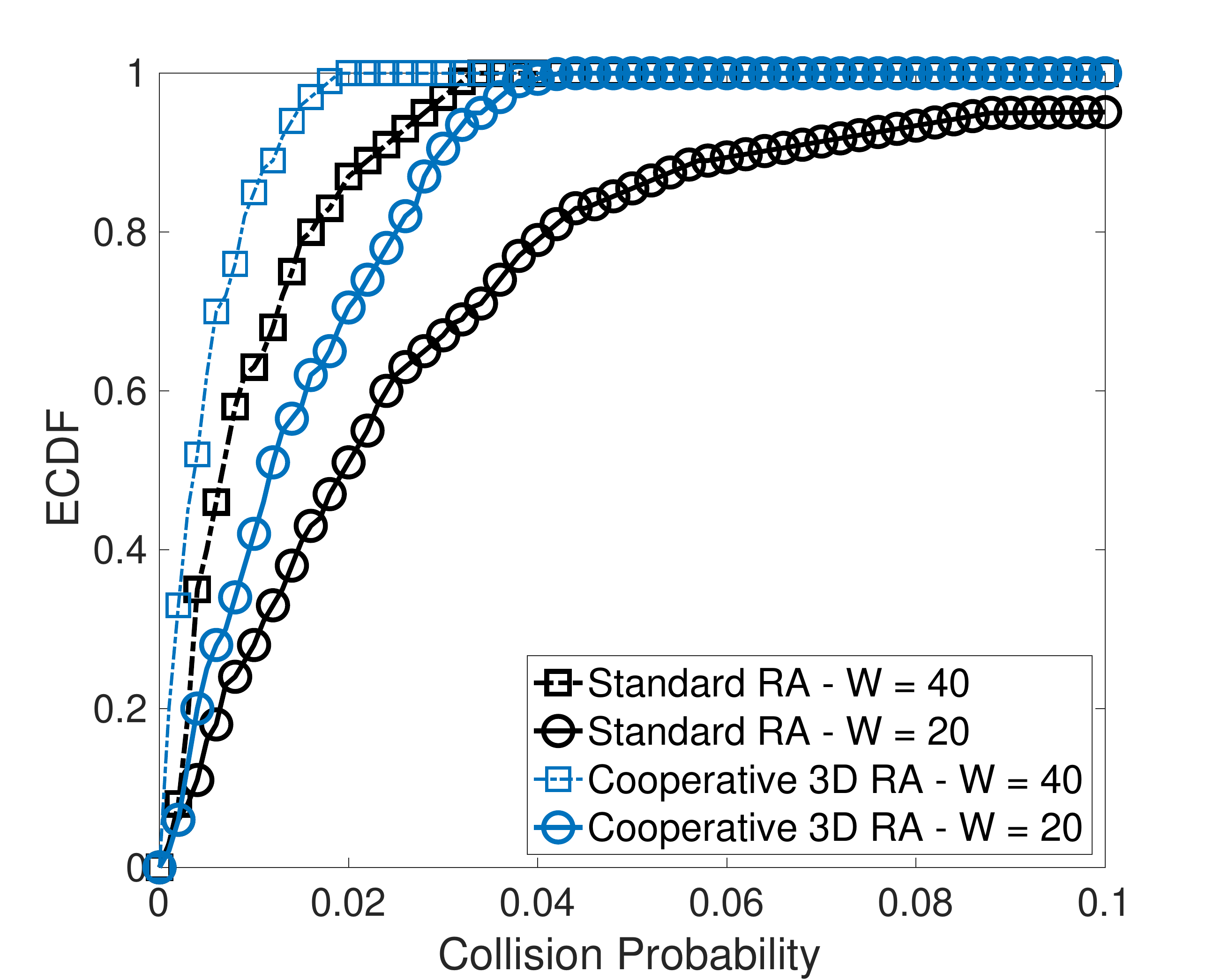}
    \caption{ECDF of the collision probability of the two RA approaches.}
    \label{fig:Coll}
\end{figure}

The performances of the RA schemes are analyzed in terms of PDR, CBR, and collision probability. The PDR is defined as
\begin{align}
PDR = \frac{\sum_{m=1}^{N_{tb}}I(i,m)}{N_{tb}},
\end{align}
where $N_{tb}$ is the number of transmitted TBs, and
\begin{align}
    I(i,m) = \begin{cases}
			1, & \text{if $i$th RxV receives the $m$th TB,}\\
            0, & \text{otherwise}.
		 \end{cases}
\end{align}

The CBR is a metric introduced by 3GPP to highlight the RP occupancy and it is defined as the ratio of busy sub-channels within a predefined interval~\cite{9345798}. % (set to 50 slots for the simulation results).
Moreover, in order to further investigate the performances of the RA schemes, the two methods are compared in terms of collision probability, defined as the probability that at least two vehicles select the same sub-channels. In the case of cooperative 3D scheme, this means that two different RxVs assign the same RA configurations to their corresponding TxVs. Assuming $N$ vehicles that simultaneously perform the selection procedure, the $i$th vehicle has a set of available sub-channels, which contains the non null elements of $\mathbf{B}_i$ in \eqref{eq:bitmap} for the standard procedure and of $\mathbf{B}_{i,\theta}$ for the cooperative 3D scheme. As the perception of an individual vehicle may differ due to the dynamic urban scenario, these sets can be distinct. 

The collision probability is defined as 
\begin{align} \label{eq:coll}
P = \frac{\sum_{m=1}^{N_{sub}}S(t,f)}{N_{sub}},
\end{align}
where $N_{sub}$ is the number of selected sub-channels, and
\begin{align}
    S(t,f) = \begin{cases}
			1, & \text{if $\geq 2$ CAVs select the $(t,f)$th sub-channel,} \\ 
            0, & \text{otherwise}.
		 \end{cases}
\end{align}
%
%
\begin{comment}
\begin{align}
    K_n =  \binom{N}{n},
\end{align}
%
with $n = 2 ... N-1$. The $n$-tuple of the colliding CAVs is referred as $\mathbf{c}_{k}$, with $k = 1 ... K_n$ and the intersection of the sets of the CAVs involved in $\mathbf{c}_{k}$ is 
%
\begin{align}
    \mathcal{I}_{n,k} = \bigcap_{i \in \mathbf{c}_{k}}\mathcal{A}_{i}
\end{align}
%
By defining the function 
%
\begin{align}
     d(n) = \begin{cases}
			1  & n = 2,\\
            -1 & \text{otherwise},
		 \end{cases}
\end{align}
%
the collision probability is
\end{comment}
%
 
\subsection{Results Discussion}
Figure~\ref{fig:PDR} depicts the PDR versus the average interference power experienced by the V2V system, obtained for a selection window of $W$ = $40$ slots and by limiting the maximum number of vehicles that can simultaneously perform the selection to $N_{max}$\,=\,$\{10,20\}$. The cooperative 3D RA shows a robust response to interference compared to the standard RA scheme, achieving on average 10\% higher PDR, being able to discriminate the occupation of resources also along the spatial dimension spanned by the beams.

The CBR of the two RA approaches are reported in Fig.~\ref{fig:CBR} in the case of a selection window $W$ = $\{20,40\}$ slots with $N_{max}$\,=\,$10$. In both scenarios, the proposed method outperforms the standard one by guaranteeing a CBR 30\% lower (obtained  by  intercepting  the  ECDF  curve at  0.5).

Figure~\ref{fig:Coll} shows that the cooperative 3D RA halves the average collision probability in \eqref{eq:coll} in the case of different $W$ and $N_{max}$ = $10$.

The cooperative 3D RA enhances the perception of the available resources by adding the third directional dimension in the BWP. This counteracts the spatial interference and improves the average PDR for the V2V system. Moreover, the CBR and collision probability results suggest that the cooperative 3D approach limits the bandwidth occupancy, although it needs a two-way communication and consequently uses more sub-channels. 

\section{Conclusion}
In mmW V2V communications, interference is a critical issue that needs to be addressed to achieve the stringent requirements of the upcoming V2X advanced services. One of the methods to reduce interference is through resource allocation schemes. In this paper, we demonstrate that the standard-complaint resource allocation procedure is not efficient in mitigating interference, as it is originally designed for the previous release, i.e., C-V2X, where beam-based communications are not contemplated. 
Motivated by this, we present a resource allocation strategy that exploits the spatial dimension spanned by the beamforming and cooperation among vehicles to mitigate V2V interference.
Extensive 3GPP-compliant numerical simulations show that the proposed approach improves the packet delivery ratio versus interference power by 10\% on average compared to the current standard method. Moreover, both channel busy ratio and collision probability are evaluated for the two methods. The performances highlight that the cooperative 3D resource allocation determines an improved perception of the available resources.

\section*{Acknowledgment}

This research was carried out in the framework of the Huawei-Politecnico di Milano Joint Research Lab activities. The Authors want to acknowledge the Huawei Milan Research Centre.

\bibliographystyle{IEEEtran}
\bibliography{biblio}

\begin{thebibliography}{10}
\providecommand{\url}[1]{#1}
\csname url@samestyle\endcsname
\providecommand{\newblock}{\relax}
\providecommand{\bibinfo}[2]{#2}
\providecommand{\BIBentrySTDinterwordspacing}{\spaceskip=0pt\relax}
\providecommand{\BIBentryALTinterwordstretchfactor}{4}
\providecommand{\BIBentryALTinterwordspacing}{\spaceskip=\fontdimen2\font plus
\BIBentryALTinterwordstretchfactor\fontdimen3\font minus
  \fontdimen4\font\relax}
\providecommand{\BIBforeignlanguage}[2]{{%
\expandafter\ifx\csname l@#1\endcsname\relax
\typeout{** WARNING: IEEEtran.bst: No hyphenation pattern has been}%
\typeout{** loaded for the language `#1'. Using the pattern for}%
\typeout{** the default language instead.}%
\else
\language=\csname l@#1\endcsname
\fi
#2}}
\providecommand{\BIBdecl}{\relax}
\BIBdecl

\bibitem{dsrc}
K.~{Abboud}, H.~A. {Omar}, and W.~{Zhuang}, ``Interworking of dsrc and cellular
  network technologies for v2x communications: A survey,'' \emph{IEEE
  Transactions on Vehicular Technology}, vol.~65, no.~12, pp. 9457--9470, 2016.

\bibitem{Sepulcre8691973}
R.~{Molina-Masegosa}, J.~{Gozalvez}, and M.~{Sepulcre}, ``Configuration of the
  c-v2x mode 4 sidelink pc5 interface for vehicular communication,'' in
  \emph{2018 14th International Conference on Mobile Ad-Hoc and Sensor Networks
  (MSN)}, 2018, pp. 43--48.

\bibitem{Sepulcre8581518}
M.~{Gonzalez-Martín}, M.~{Sepulcre}, R.~{Molina-Masegosa}, and J.~{Gozalvez},
  ``Analytical models of the performance of c-v2x mode 4 vehicular
  communications,'' \emph{IEEE Transactions on Vehicular Technology}, vol.~68,
  no.~2, pp. 1155--1166, 2019.

\bibitem{9127428}
F.~Tang, Y.~Zhou, and N.~Kato, ``Deep reinforcement learning for dynamic
  uplink/downlink resource allocation in high mobility 5g hetnet,'' \emph{IEEE
  Journal on Selected Areas in Communications}, vol.~38, no.~12, pp.
  2773--2782, 2020.

\bibitem{he2020interference}
X.~He, J.~Lv, J.~Zhao, X.~Hou, and T.~Luo, ``Design and analysis of a
  short-term sensing-based resource selection scheme for c-v2x networks,''
  \emph{IEEE Internet of Things Journal}, vol.~7, no.~11, pp. 11\,209--11\,222,
  2020.

\bibitem{molina2019geo}
R.~Molina-Masegosa, M.~Sepulcre, and J.~Gozalvez, ``Geo-based scheduling for
  c-v2x networks,'' \emph{IEEE Transactions on Vehicular Technology}, vol.~68,
  no.~9, pp. 8397--8407, 2019.

\bibitem{kim2018position}
J.~Kim, J.~Lee, S.~Moon, and I.~Hwang, ``A position-based resource allocation
  scheme for v2v communication,'' \emph{Wireless Personal Communications},
  vol.~98, no.~1, pp. 1569--1586, 2018.

\bibitem{zhao2019cluster}
J.~Zhao, X.~He, H.~Wang, X.~Zheng, J.~Lv, T.~Luo, and X.~Hou, ``Cluster-based
  resource selection scheme for 5g v2x,'' in \emph{2019 IEEE 89th Vehicular
  Technology Conference (VTC2019-Spring)}.\hskip 1em plus 0.5em minus
  0.4em\relax IEEE, 2019, pp. 1--5.

\bibitem{6g}
K.~Sakaguchi, R.~Fukatsu, T.~Yu, E.~Fukuda, K.~Mahler, R.~Heath, T.~Fujii,
  K.~Takahashi, A.~Khoryaev, S.~Nagata \emph{et~al.}, ``Towards mmwave v2x in
  5g and beyond to support automated driving,'' \emph{IEICE Transactions on
  Communications}, 2020.

\bibitem{Rappaport:J13}
T.~S. {Rappaport}, S.~{Sun}, R.~{Mayzus}, H.~{Zhao}, Y.~{Azar}, K.~{Wang},
  G.~N. {Wong}, J.~K. {Schulz}, M.~{Samimi}, and F.~{Gutierrez}, ``{Millimeter
  Wave Mobile Communications for 5G Cellular: It Will Work!}'' \emph{IEEE
  Access}, vol.~1, pp. 335--349, May 2013.

\bibitem{Petrov_inter}
V.~Petrov, J.~Kokkoniemi, D.~Moltchanov, J.~Lehtomäki, M.~Juntti, and
  Y.~Koucheryavy, ``The impact of interference from the side lanes on
  mmwave/thz band v2v communication systems with directional antennas,''
  \emph{IEEE Transactions on Vehicular Technology}, vol.~67, no.~6, pp.
  5028--5041, 2018.

\bibitem{TS_38214}
{3GPP TS 38.214 V16.3.0}, ``{3rd Generation Partnership Project; Technical
  Specification Group Radio Access Network; NR; Physical layer procedures for
  data (Release 16)},'' Sep. 2020.

\bibitem{sidelink}
S.~{Lien}, D.~{Deng}, C.~{Lin}, H.~{Tsai}, T.~{Chen}, C.~{Guo}, and S.~{Cheng},
  ``3gpp nr sidelink transmissions toward 5g v2x,'' \emph{IEEE Access}, vol.~8,
  pp. 35\,368--35\,382, 2020.

\bibitem{LinsICC-c21}
F.~Morandi, F.~Linsalata, M.~Brambilla, M.~Mizmizi, M.~Magarini, and
  U.~Spagnolini, ``A probabilistic codebook technique for fast initial access
  in {6G} {Vehicle-to-Vehicle} communications,'' in \emph{IEEE Int. Conf.
  Commun. Workshops}, 2021.

\bibitem{phychanmod}
{3GPP 38.211 V16.0.0}, ``3rd generation partnership project; technical
  specification nr; physical channels and modulation (release 16),'' Mar. 2020.

\bibitem{mode1}
{LG Electronics}, ``Summary of ran1 agreements/working assumptions in wi 5g v2x
  with nr sidelink,'' 3GPP TSG RAN WG1 Meeting 99, Tech. Rep. R1-1913601, Nov.
  2019, Reno, USA.

\bibitem{9345798}
M.~H.~C. Garcia, A.~Molina-Galan, M.~Boban, J.~Gozalvez, B.~Coll-Perales,
  T.~Şahin, and A.~Kousaridas, ``A tutorial on 5g nr v2x communications,''
  \emph{IEEE Communications Surveys Tutorials}, vol.~23, no.~3, pp. 1972--2026,
  2021.

\bibitem{garcia2021tutorial}
M.~H.~C. Garcia, A.~Molina-Galan, M.~Boban, J.~Gozalvez, B.~Coll-Perales,
  T.~{\c{S}}ahin, and A.~Kousaridas, ``A tutorial on 5g nr v2x
  communications,'' \emph{arXiv preprint arXiv:2102.04538}, 2021.

\bibitem{8613007}
N.~Bonjorn, F.~Foukalas, F.~Cañellas, and P.~Pop, ``Cooperative resource
  allocation and scheduling for 5g ev2x services,'' \emph{IEEE Access}, vol.~7,
  pp. 58\,212--58\,220, 2019.

\bibitem{bazzi2018study}
A.~Bazzi, G.~Cecchini, A.~Zanella, and B.~M. Masini, ``Study of the impact of
  phy and mac parameters in 3gpp c-v2v mode 4,'' \emph{IEEE Access}, vol.~6,
  pp. 71\,685--71\,698, 2018.

\bibitem{15rel}
{3GPP TR 37.985}, ``Technical specification group radio access network; overall
  description of radio access network (ran) aspects for vehicle-to-everything
  (v2x) based on lte and nr,'' 2020.

\bibitem{TS_38213}
{3GPP TS 38.213 V16.3.0}, ``{3rd Generation Partnership Project; Technical
  Specification Group Radio Access Network; NR; Physical layer procedures for
  control (Release 16)},'' Sep. 2020.

\bibitem{14rel}
{3GPP TR 37.885 v15.3.0}, ``{Study on evaluation methodology of new
  Vehicle-to-Everything (V2X) use cases for LTE and NR (Release 15)},'' Jun.
  2019.

\bibitem{Perfecto:J2017}
C.~{Perfecto}, J.~{Del Ser}, and M.~{Bennis}, ``{Millimeter-Wave V2V
  Communications: Distributed Association and Beam Alignment},'' \emph{IEEE J.
  Sel. Areas Commun.}, vol.~35, no.~9, pp. 2148--2162, Jun. 2017.

\bibitem{osm}
{OpenStreetMap contributors}, ``{Planet dump retrieved from
  https://planet.osm.org },'' \url{ https://www.openstreetmap.org}, 2017.

\bibitem{SUMO2018}
\BIBentryALTinterwordspacing
P.~A. {Lopez et al.}, ``{Microscopic Traffic Simulation using SUMO},'' in
  \emph{IEEE Int. Conf. Intell. Transp. Syst.}\hskip 1em plus 0.5em minus
  0.4em\relax IEEE, Dec. 2018. [Online]. Available:
  \url{https://elib.dlr.de/124092/}
\BIBentrySTDinterwordspacing

\bibitem{GEMV2}
M.~{Boban}, J.~{Barros}, and O.~K. {Tonguz}, ``{Geometry-Based
  Vehicle-to-Vehicle Channel Modeling for Large-Scale Simulation},'' \emph{IEEE
  Transaction Vehicular Technology}, vol.~63, no.~9, pp. 4146--4164, Apr. 2014.

\end{thebibliography}
\end{document}